\documentclass[10pt,conference]{IEEEtran}
\IEEEoverridecommandlockouts
\usepackage{url}
\usepackage{cite}
\usepackage{amsmath,amssymb,amsfonts}
\usepackage{algorithmic}
\usepackage{graphicx}
\usepackage{svg}
\usepackage{textcomp}
\usepackage{xcolor}
\usepackage{xspace}
\usepackage{pgfplots}
\usepackage{tikz}
\usepackage{algorithm}
\usepackage{algorithmic}
\usepackage[T1]{fontenc}
\usepackage{multirow}
\usepackage{lscape}
\usepackage{subcaption}
\def\BibTeX{{\rm B\kern-.05em{\sc i\kern-.025em b}\kern-.08em
    T\kern-.1667em\lower.7ex\hbox{E}\kern-.125emX}}
    
\newcommand{\NBZ}{\mathcal{Z}\xspace} 
\newcommand{\NbNoeudsCentraux}{\mathcal{N}\xspace}

\newcommand{\metriconesymbol}{m_1}
\newcommand{\metrictwosymbol}{m_2}
\newcommand{\metricthreesymbol}{m_3}
\newcommand{\metricfoursymbol}{m_4}

\begin{document}

\title{Tool Demo: Topology analysis with GPML for detection of cyberattacks in Water Distribution Networks (Invited demo)\\
\thanks{This study was funded by French ANR under grant ANR-22-CE39-0010 for CoRREau Project.}
}

\author{\IEEEauthorblockN{Majed JABER}
\IEEEauthorblockA{\textit{ICube UMR7357} \\
\textit{Université de Strasbourg}\\
67000 Strasbourg, France \\
jaberm@unistra.fr}
\and
\IEEEauthorblockN{Abdul Qadir KHAN}
\IEEEauthorblockA{\textit{Laboratoire de Recherche de l'EPITA} \\
Le Kremlin-Bic\^{e}tre, France \\
abdul-qadir.khan@epita.fr}
\and
\IEEEauthorblockN{Ankush MESHRAM}
\IEEEauthorblockA{\textit{KASTEL Security Research Labs} \\ \textit{Vision and Fusion Laboratory (IES)} \\
\textit{Karlsruhe Institute of Technology}\\
Karlsruhe, Germany \\
ankush.meshram@kit.edu}
\and
\IEEEauthorblockN{Julien MICHEL}
\IEEEauthorblockA{\textit{Laboratoire de Recherche de l'EPITA} \\
Le Kremlin-Bic\^{e}tre, France \\
julien.michel@epita.fr}
\and
\IEEEauthorblockN{Côme FRAPPE-VIALATOUX}
\IEEEauthorblockA{\textit{Laboratoire de Recherche de l'EPITA} \\
Le Kremlin-Bic\^{e}tre, France \\
come.frappe-vialatoux@epita.fr}
\and
\IEEEauthorblockN{Pierre PARREND}
\IEEEauthorblockA{\textit{Laboratoire de Recherche de l'EPITA}\\ 
Le Kremlin-Bic\^{e}tre, France \\
\textit{Université de Strasbourg, CNRS, ICube}\\ 
UMR 7357, 67000 Strasbourg, France \\
pierre.parrend@epita.fr}
}

\maketitle

\begin{abstract}
Water distribution networks depends on industrial control systems to integrate the physical process with communication network, making them vulnerable to cyberattacks that alter the traffic pattern and network behavior. Traditional detection approaches that rely on raw traffic or protocol information often oversee structural changes that are induced by such attacks. In this work, we present a topology-driven approach for detection of cyberattacks in water distribution networks based on the Graph Processing for Machine Learning (GPML) framework. The raw traffic is transformed into dynamic graphs, from which community and spectral metrics are extracted and analyzed for any structural and communication modifications with time. The proposed methodology is evaluated on three industrial water distribution datasets such as HITL, SWaT, and CrossTest. Spectral and community graph metrics improve the model performance in detection of cyber and physical attacks across the three datasets. 
\end{abstract}


\section{Introduction}
Water Distribution Networks (WDNs) constitute large-scale systems that integrate physical infrastructure with Industrial Control Systems (ICS) and digital communication networks for monitoring, control, and automation. Sensors, programmable logic controllers, and supervisory platforms exchange continuous streams of heterogeneous operational data over communication protocols, forming complex networked environments where physical processes and traffic data are tightly coupled. This convergence exposes WDNs to variety of cyber threats that target both logical communication layers and physical control processes. Network-level attacks such as Man-in-the-Middle (MitM), traffic scanning, and reconnaissance exploit communication channels to intercept, alter, or infer control flows, while physical-level attacks impact system dynamics and physical safety. These attacks often can be treated as structural and topological anomalies in network traffic, rather than explicit signature patterns, making detection increasingly dependent on behavioral and topology-driven analysis. In this context, modeling WDN communication flows as dynamic networks enables the extraction of structural, spectral, and community-level properties that reflect system organization, connectivity, and interaction patterns. Topology analysis therefore provides a principled framework for detecting cyberattacks through deviations in connectivity structure, information flow, and spectral characteristics of traffic graphs. This paper presents a graph-based topology-driven detection approach through Graph Processing for Machine Learning (GPML) library~\cite{jaber2025gpml}, through community and spectral indicators to identify anomalous behaviors in WDN network traffic. By leveraging graph representations and spectral signatures, the proposed framework captures both local interaction anomalies and global structural disruptions induced by cyberattacks, enabling interpretable and scalable detection mechanisms for industrial water infrastructure.


\section{Topology Analysis for Attack Detection}
\label{sec.topology}
Various methods in the literature rely on the logs or raw traffic data and use these logs for monitoring the activities in a network. These logs provides valuable information but some hidden attack patterns can go unseen~\cite{shaukat2025}. Therefore, a topology based analysis where these logs are transformed into a graph based representation which provides an overview of the structure and the communication in the network can help detect these anomalous patterns. In such topology based representation, the nodes are the devices such as computer, servers, switches or routers, and the communication between these nodes is represented as edges. 

Once the network is represented in the form of graphs, it becomes possible to analyze complex topology and interactions, and to detect irregularities using graph-based methods. More specifically, in cybersecurity perspective, the analysis of such graphs can help in understanding the evolution of the network over time and identify anomalies. In such a case, any significant change in the topology indicates a cyberattack on the network. Graph spectral and community metrics are the metrics that can be used to analyze the network graphs. These metrics provide structural properties of evolving graph with time. These metrics are explained in detail in Section~\ref{sec: gpml}. 

\section{Spectral and Topology-Based Intrusion Detection in ICS}

Recent research increasingly investigates graph-based and spectral approaches for detecting cyberattacks in industrial control systems (ICS). Unlike signature-driven monitoring of raw traffic logs, topology-aware methods model communication patterns as dynamic graphs, enabling the analysis of structural deviations induced by attacks.

\textbf{Spectral change detection.}
Spectral analysis of evolving network graphs has been used to capture global structural disruptions such as denial-of-service propagation, lateral movement, and coordinated scanning. Early work on network-wide traffic anomalies demonstrated that eigenvalue variations can reveal macroscopic changes in communication flows \cite{lakhina2004characterization}. More recent studies extend spectral monitoring to cyber-physical systems, showing that Laplacian spectrum indicators can characterize connectivity fragmentation and traffic concentration under attack conditions
\cite{jaber2024spectral}.

\textbf{Topology-based intrusion detection in ICS.}
Industrial networks present deterministic communication patterns, which makes topology-driven anomaly detection particularly relevant. Graph modeling of PLC–SCADA interactions enables detection of deviations in control loops and communication paths. Testbeds such as SWaT and cross-domain ICS environments have been widely adopted to evaluate such approaches \cite{karch2022crosstest}. Recent surveys highlight the importance of multi-step attack modeling and graph representations to improve detection robustness in ICS environments \cite{shaukat2025review}.

\section{Graph Processing for Machine Learning}
\label{sec: gpml}
The GPML\footnote{\url{https://github.com/lre-security-systems-team/gpml}} framework was previously introduced~\cite{jaber2025cyberattack, jaber2025gpml} as a graph-based cybersecurity analysis library that transforms raw network traffic traces into structured graph representations. GPML enables the analysis of dynamic network behavior through \textbf{community metrics} and \textbf{spectral metrics}, supporting detection of different types of attacks. These added metrics can detect focus on monitoring structural shifts, and evolving communication patterns. The library supports forensic analysis by modeling traffic as dynamic graphs and extracting interpretable features for investigation workflows.

\subsection{Community Metrics}
\begin{figure}[h]
    \centering
    \begin{minipage}{0.5\textwidth}
        \centering
        \includegraphics[width=\textwidth]{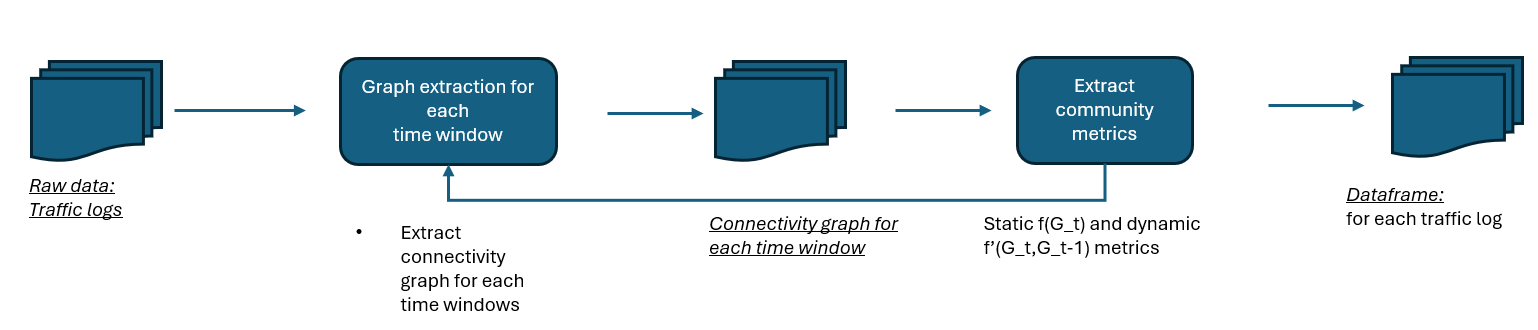}
        \caption{Flow process diagram of the community metrics methodology.}
        \label{fig.community_pipeline}
    \end{minipage}
\end{figure}

The graph community metrics are computed using a pipeline shown in Fig.~\ref{fig.community_pipeline}. Community metrics represents the network topology at a given time, hence, these metrics can be used to detect different attacks in a network. First two types of graphs for a time window are extracted from the raw data. IP graphs where the nodes represents the IP addresses and the communication between these devices is represented as edges, while, the second graph is an extension of the IP graph where nodes are represented as a pair of IP and Port addresses. From this, we obtain a connectivity graph that show the network topology and communication at a selected time window. In our experiments we choose the time window of 5 seconds. Now to extract the metrics from the connectivity graphs, we use the Louvain algorithm~\cite{louvain} to partition these graphs into communities. We calculate the graph community metrics from the partition at time $t$ and their dynamicity at time $t+1$. Stability, Density, Conductance, and Degree are extracted. To define these metrics, let $V_{t}$ be the set of nodes of a community at time $t$. 

\noindent\textbf{Stability} Measure the similarity between the states of community at time $t$ and $t+1$.
 \[
        Stability =
        \frac{|V_t \cap V_{t+1}| - |(V_t \cap \bar{V}_{t+1}) \cup (V_{t+1} \cap \bar{V}_t)|}
        {|V_t \cup V_{t+1}|}.
        \]
\noindent \textbf{Density} Calculates the probability of the node being adjacent to another node within the community. 

\noindent \textbf{Conductance} Measures the fraction of communications that link nodes inside the community to nodes outside it.

\noindent \textbf{Degree} Measures the number of edges going out of a node in a community. 

\subsection{Spectral Metrics}

\begin{figure}[h]
    \centering
    \begin{minipage}{0.5\textwidth}
        \centering
        \includegraphics[width=\textwidth]{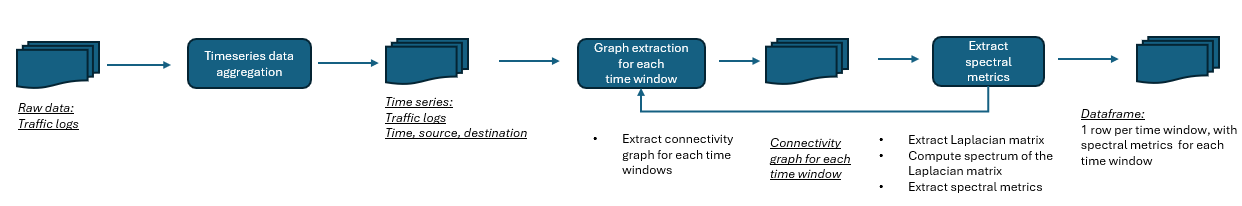}
        \caption{Flow process diagram of the dynamic spectral metrics methodology.}
        \label{fig.spectral_pipeline}
    \end{minipage}
\end{figure}

We compute spectral indicators within a progressive pipeline that moves from time-series aggregation to time-windowing as shown in Fig.~\ref{fig.spectral_pipeline}, then to spectral time-windowing (you can find the detailed methodology in~\cite{jaber2024graph}). First, we transform raw traffic logs into a time-series representation by aggregating records that share the same communication context (e.g., same source and destination) within a fixed interval~\cite{lakhina2004characterization}. 

This step reduces log imbalance, regularizes the temporal axis, and produces stable quantitative features (packets, bytes, rates) per timestamp. Second, we apply time-windowing by sliding a window $tw(t)$ over the time-series to capture short-term temporal evolution~\cite{lin2019anomaly}. Each window aggregates the quantitative features inside the window and produces one row per window, which enables window-level detection. Third, we extend time-windowing with topology: for each time-window, we build weighted graphs from the communications observed inside the window (nodes represent entities and weighted edges represent interaction intensity such as packets, bytes, or rates). We then compute the Laplacian spectrum $\Lambda(t)$ of each graph to extract compact indicators that track how connectivity and weight distribution evolve over time. Let $G_t$ denote the graph built at time $t$, $L_t$ its Laplacian, and $\Lambda(t)=\{\Lambda(t)[1],\dots,\Lambda(t)[n]\}$ the eigenvalues sorted in increasing order. Let $\NBZ(t)$ be the multiplicity of zero in $\Lambda(t)$ (number of connected components), and let $\NbNoeudsCentraux$ denote the number of central devices (e.g., servers or switches) used to parameterize eigenvalue ranges. We define four metrics that map spectral properties to interpretable network dynamics: \textbf{Connectedness} measures interconnectivity through component reduction and is quantified from the number of connected components $\NBZ(t)$ as $\metriconesymbol(t)=\exp\!\big(\tfrac{1}{\NBZ(t)}\big)/\exp(1)$. 
\textbf{Flooding} captures weight-driven strengthening in the first non-zero eigenvalue band by averaging the $\NbNoeudsCentraux$ eigenvalues immediately after the zero block as $\metrictwosymbol(t)=\left(\tfrac{1}{\NbNoeudsCentraux}\sum_{i=\NBZ(t)+1}^{\NBZ(t)+\NbNoeudsCentraux}\lambda_i^{(t)}\right)-1$. 
\textbf{Wiriness} reflects activity concentration in the largest eigenvalues by averaging the $\NbNoeudsCentraux$ highest spectral values as $\metricthreesymbol(t)=\tfrac{1}{\NbNoeudsCentraux}\sum_{i=n-\NbNoeudsCentraux+1}^{n}\lambda_i^{(t)}$. 
\textbf{Asymmetry} measures spectral irregularity through eigen-gap counting as $\metricfoursymbol(t)=\#\{k\in[2,n]\;|\;\Lambda(t)[k]-\Lambda(t)[k-1]>\varepsilon\}$ with $\varepsilon=10^{-12}$.


\section{Attack detection in Industrial Control Systems for Water distribution}
\label{sec:wdsICS}
\begin{figure}[h]
    \centering
    \begin{minipage}{0.4\textwidth}
        \centering
        \includegraphics[width=\textwidth]{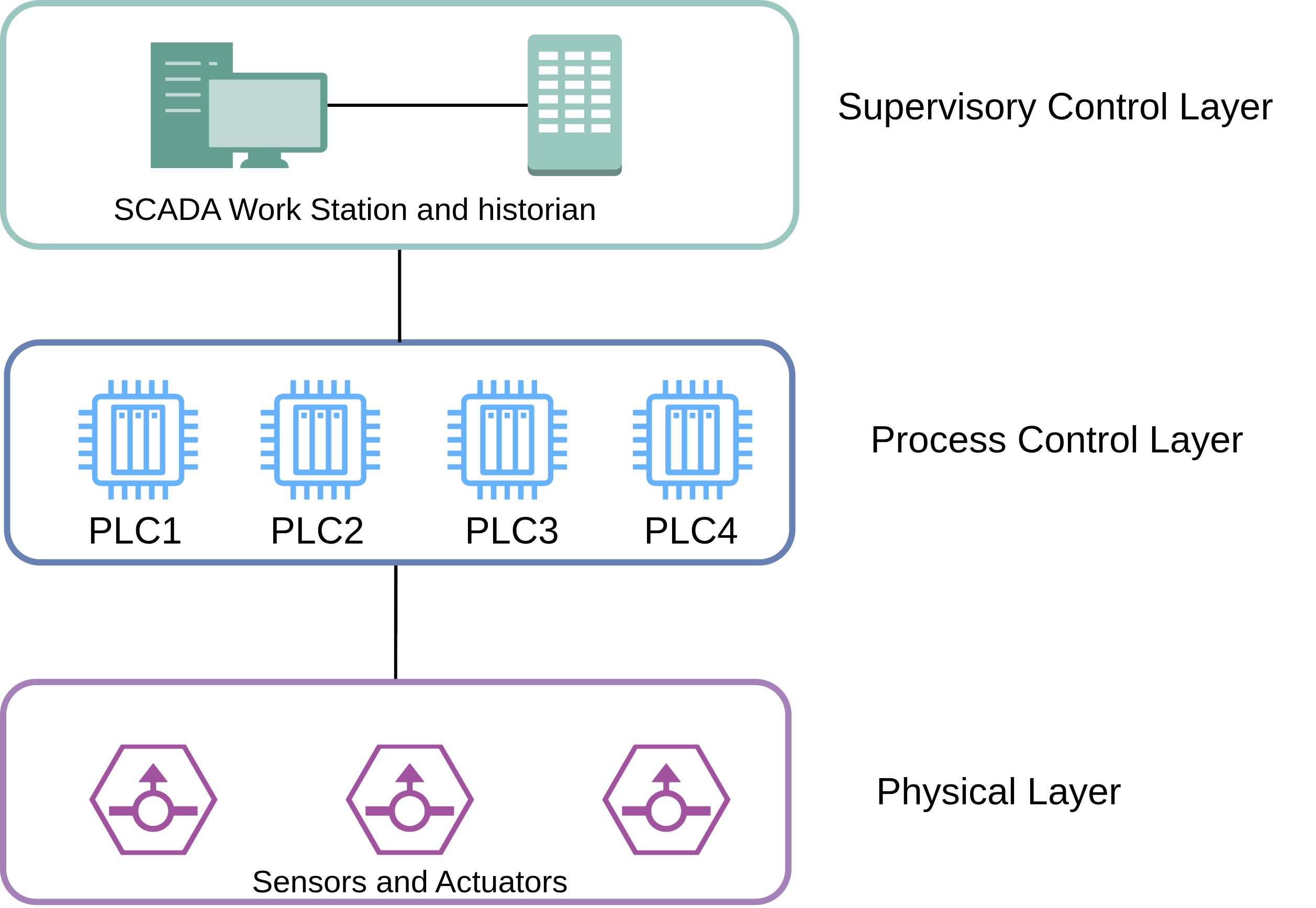}
        \caption{Architecture of Water Distribution ICS}
        \label{fig.hitlnetwork}
    \end{minipage}
\end{figure}

In this  Section, we present an overview of the three ICSs for water distribution that were used in our experiments. An overview of the water distribution architecture is shown in Fig.~\ref{fig.hitlnetwork}. A three layered architecture has been widely used for water distribution. The \textbf{Supervisory Control Layer} consists of Supervisory Control And Data Acquisition (SCADA) server system, Human Machine Interface (HMI) console, and Historian. HMI allows the end administrator or controller to have access to the server and the records stored in the historian. SCADA is used as a main supervisory node to oversee the whole process of water distribution. The \textbf{Process Control Layer} consists of multiple Programmable Logic Controllers (PLCs) that collect data from the sensors and actuators of the Physical layer and provide information to the Supervisory layer. The \textbf{Physical Layer} consists of physical objects that are deployed at the ground level such as sensors and actuators, pumps, tanks, valves, and reservoirs. 


Hardware in the loop (HITL) is a dataset derived from water distribution testbed that consists of the hardware and simulated parts~\cite{Faramondi2021Hitl}. The integration of physical and simulated components results in a sophisticated system that allows the exploration of a wide range of cyber-physical attacks. This dataset comprises physical data obtained from a programmable logic controller (PLC), as well as the network data captured from the communication in the running testbed. There are two categories of attacks considered that are physical attacks and cyber attacks. Physical attacks are related to malfunction of sensors, valves, or pumps while cyber attacks are MiTM, Denial of Service (DoS), and scanning attacks. 

Secure Water Treatment (SWaT) is a dataset published by iTrust research center at Singapore University of Technology and Design~\cite{Mathur2016Swat}. The testbed is used to generate different datasets where different attacks has been simulated. There are multiple updated datasets provided since 2016 and has been used widely in the literature for detection of attacks in water distribution system. The one we used in our experiments is SWat.A6 2019 due to the presence of the different attacks in the dataset. 

\begin{figure}[h]
    \centering
    \begin{minipage}{0.4\textwidth}
        \centering
        \includegraphics[width=\textwidth]{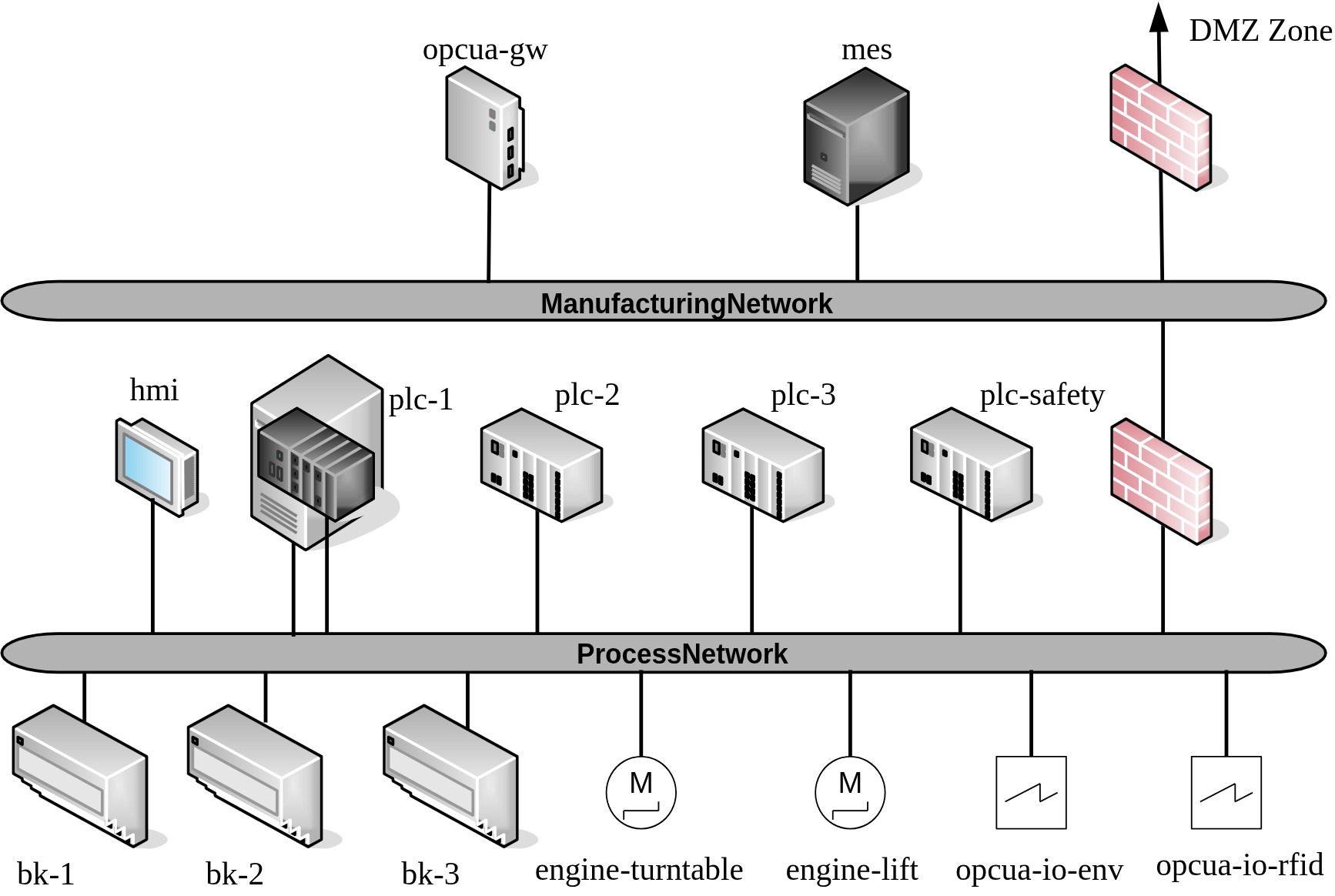}
        \caption{Architecture of CrossTest Platform}
        \label{fig.crosstest}
    \end{minipage}
\end{figure}

CrossTest\footnote{\url{https://fordatis.fraunhofer.de/handle/fordatis/314}} is a cross-domain physical cybersecurity testbed environment for the development and evaluation of domain-agnostic threat detection methods~\cite{crosstest}. The overview of the architecture of the testbed is provided in Fig~\ref{fig.crosstest}. Multiple attacks were implemented on Energy and Production domain testbeds, the corresponding network traffic is captured and made available as PCAP files. For the presented demo, we utilize the data captured from Production testbed containing a kill chain of Scanning, Automated Collection and Denial-of-Service Profinet attacks in the order, along with normal operation.


\section{Evaluations}
\label{sec:evaluations}

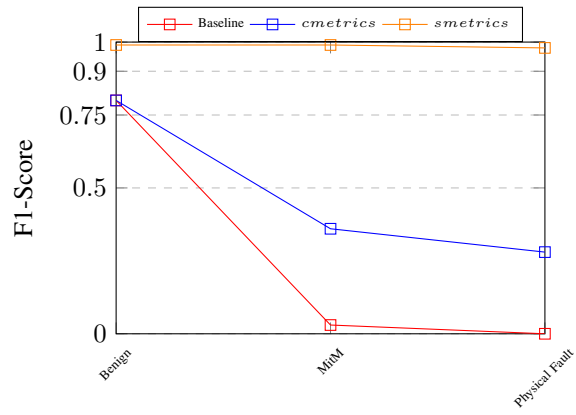
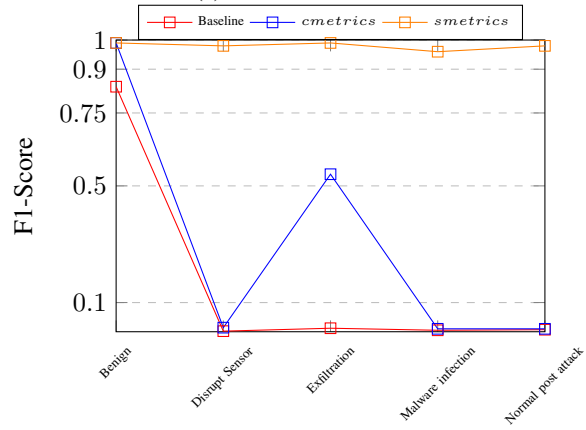
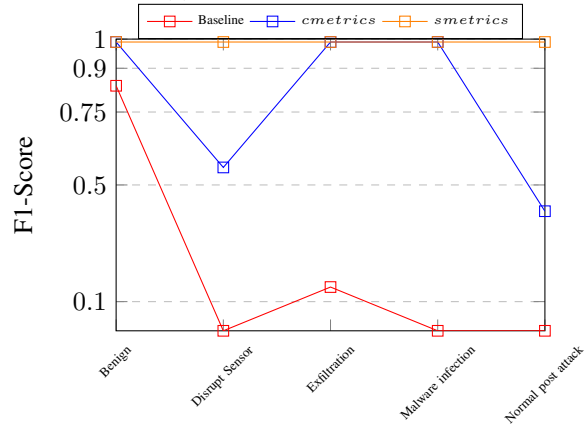
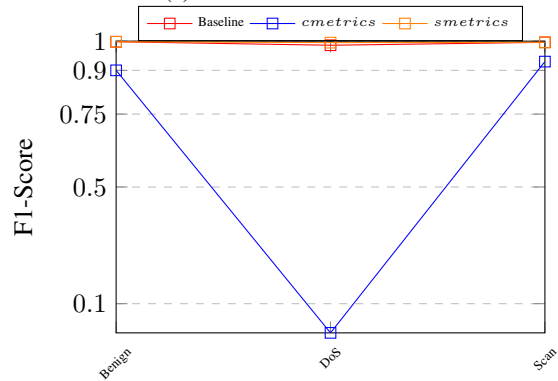
\begin{figure}[ht]
\centering
\begin{subfigure}[b]{0.4\textwidth}
    \centering
    \begin{tikzpicture}
    \begin{axis}[
        ylabel={F1-Score},
        xmin=0, xmax=2, ymin=0.0, ymax=1,
        xtick={0,1,2,3}, xticklabels={Benign, MitM, Physical Fault},
        xticklabel style={font=\tiny,rotate=45},
        ytick={0.0, 0.5,0.75,0.9,1}, legend pos=south west,
        legend style={fill=none, font=\tiny}, 
        ymajorgrids=true, grid style=dashed,
        width=\textwidth, height=0.75\textwidth,
        legend style={
        at={(0.5,1.0)},
        anchor=south,
        legend columns=3
        }
    ]
    \addplot[color=red,mark=square] coordinates {
        (0,0.80)(1,0.03)(2,0.00)};
    \addlegendentry{Baseline}
    \addplot[color=blue,mark=square] coordinates {
        (0,0.80)(1,0.36)(2,0.28)};
    \addlegendentry{$cmetrics$}
    \addplot[color=orange,mark=square] coordinates {
        (0,0.99)(1,0.99)(2,0.98)};
    \addlegendentry{$smetrics$}
    \end{axis}
    \end{tikzpicture}
    \caption{HiTL Dataset}
    \label{fig:bot.f1score.9}
\end{subfigure}
\hspace{0.02\textwidth}
\begin{subfigure}[b]{0.4\textwidth}
    \centering
    \begin{tikzpicture}
    \begin{axis}[
        ylabel={F1-Score},
        xmin=0, xmax=4, ymin=0.0, ymax=1,
        xtick={0,1,2,3,4}, xticklabels={Benign, Disrupt Sensor, Exfiltration, Malware infection, Normal post attack},
        xticklabel style={font=\tiny,rotate=45},
        ytick={0.1,0.5,0.75,0.9,1}, legend pos=south west,
        legend style={fill=none, font=\tiny}, 
        ymajorgrids=true, grid style=dashed,
        width=\textwidth, height=0.75\textwidth,
        legend style={
        at={(0.5,1.0)},
        anchor=south,
        legend columns=3
        }
    ]
    \addplot[color=red,mark=square] coordinates {
        (0,0.84)(1,0.002)(2,0.012)(3,0.005)(4,0.007)};
    \addlegendentry{Baseline}
    \addplot[color=blue,mark=square] coordinates {
        (0,0.99)(1,0.014)(2,0.54)(3,0.01)(4,0.01)};
    \addlegendentry{$cmetrics$}
    \addplot[color=orange,mark=square] coordinates {
        (0,0.99)(1,0.98)(2,0.99)(3,0.96)(4,0.98)};
    \addlegendentry{$smetrics$}
    \end{axis}
    \end{tikzpicture}
    \caption{SWaT 100k Dataset}
    \label{fig:bot.f1score.10}
\end{subfigure}
\hspace{0.02\textwidth}
\begin{subfigure}[b]{0.4\textwidth}
    \centering
    \begin{tikzpicture}
    \begin{axis}[
        ylabel={F1-Score},
        xmin=0, xmax=4, ymin=0.0, ymax=1,
        xtick={0,1,2,3,4}, xticklabels={Benign, Disrupt Sensor, Exfiltration, Malware infection, Normal post attack},
        xticklabel style={font=\tiny,rotate=45},
        ytick={0.1,0.5,0.75,0.9,1}, legend pos=south west,
        legend style={fill=none, font=\tiny}, 
        ymajorgrids=true, grid style=dashed,
        width=\textwidth, height=0.75\textwidth,
        legend style={
        at={(0.5,1.0)},
        anchor=south,
        legend columns=3
        }
    ]
    \addplot[color=red,mark=square] coordinates {
        (0,0.84)(1,0.00)(2,0.15)(3,0.00)(4,0.00)};
    \addlegendentry{Baseline}
    \addplot[color=blue,mark=square] coordinates {
        (0,0.99)(1,0.56)(2,0.99)(3,0.99)(4,0.41)};
    \addlegendentry{$cmetrics$}
    \addplot[color=orange,mark=square] coordinates {
        (0,0.99)(1,0.99)(2,0.99)(3,0.99)(4,0.99)};
    \addlegendentry{$smetrics$}
    \end{axis}
    \end{tikzpicture}
    \caption{SWaT 500k Dataset}
    \label{fig:bot.f1score.11}
\end{subfigure}
\hspace{0.02\textwidth}
\begin{subfigure}[b]{0.4\textwidth}
    \centering
    \begin{tikzpicture}
    \begin{axis}[
        ylabel={F1-Score},
        xmin=0, xmax=2, ymin=0.0, ymax=1,
        xtick={0,1,2}, xticklabels={Benign, DoS, Scan},
        xticklabel style={font=\tiny,rotate=45},
        ytick={0.1,0.5,0.75,0.9,1}, legend pos=south west,
        legend style={fill=none, font=\tiny}, 
        ymajorgrids=true, grid style=dashed,
        width=\textwidth, height=0.75\textwidth,
        legend style={
        at={(0.5,1.0)},
        anchor=south,
        legend columns=3
        }
    ]
    \addplot[color=red,mark=square] coordinates {
        (0,0.998)(1,0.986)(2,0.996)};
    \addlegendentry{Baseline}
    \addplot[color=blue,mark=square] coordinates {
        (0,0.90)(1,0.00)(2,0.93)};
    \addlegendentry{$cmetrics$}
    \addplot[color=orange,mark=square] coordinates {
        (0,0.998)(1,0.995)(2,0.996)};
    \addlegendentry{$smetrics$}
    \end{axis}
    \end{tikzpicture}
    \caption{CrossTest Dataset}
    \label{fig:bot.f1score.12}
\end{subfigure}
\caption{Class-wise F1-Score results for three approaches across the HiTL, SWaT 100k, SWaT 500k, and CrossTest datasets.}
\label{fig:model_comparison:f1score}
\end{figure}

\begin{figure}[ht]
\centering

\begin{subfigure}[b]{\columnwidth}
\centering
\begin{tikzpicture}
\begin{axis}[
    ylabel={Balanced Accuracy},
    ymin=0, ymax=1,
    ymajorgrids=true, grid style=dashed,
    width=\columnwidth, height=0.6\textwidth,
    symbolic x coords={HiTL,SWaT 100k,SWaT 500k,CrossTest},
    xtick=data,
    xticklabel style={font=\small,rotate=30,anchor=east},
    bar width=6pt,
    ytick={0,0.25,0.5,0.75,0.9,1},
    legend pos=south west,
    legend style={fill=none, font=\small},
    enlarge x limits=0.18,
    ybar,
    legend style={
    at={(0.5,1.0)},
    anchor=south,
    legend columns=3
    }
]
\addplot coordinates {
    (HiTL,0.33) (SWaT 100k,0.20) (SWaT 500k,0.21) (CrossTest,0.98)
};
\addlegendentry{Baseline}

\addplot coordinates {
    (HiTL,0.47) (SWaT 100k,0.40) (SWaT 500k,0.80) (CrossTest,0.47)
};
\addlegendentry{Community}

\addplot coordinates {
    (HiTL,0.99) (SWaT 100k,0.98) (SWaT 500k,0.99) (CrossTest,0.99)
};
\addlegendentry{Spectral}

\end{axis}
\end{tikzpicture}
\caption{Balanced Accuracy comparison across datasets using Baseline, Community, and Spectral methods.}

\end{subfigure}

\begin{subfigure}[b]{\columnwidth}
\centering
\begin{tikzpicture}
\begin{axis}[
    ylabel={MCC},
    ymin=0, ymax=1,
    ymajorgrids=true, grid style=dashed,
    width=\columnwidth, height=0.55\textwidth,
    symbolic x coords={HiTL,SWaT 100k,SWaT 500k,CrossTest},
    xtick=data,
    xticklabel style={font=\small,rotate=30,anchor=east},
    bar width=6pt,
    ytick={0,0.25,0.5,0.75,0.9,1},
    legend pos=south west,
    legend style={fill=none, font=\small},
    enlarge x limits=0.18,
    ybar,
    legend style={
    at={(0.5,1.0)},
    anchor=south,
    legend columns=3
    }
]
\addplot coordinates {
    (HiTL,0.10) (SWaT 100k,0.10) (SWaT 500k,0.15) (CrossTest,0.99)
};
\addlegendentry{Baseline}

\addplot coordinates {
    (HiTL,0.28) (SWaT 100k,0.65) (SWaT 500k,0.85) (CrossTest,0.77)
};
\addlegendentry{Community}

\addplot coordinates {
    (HiTL,0.98) (SWaT 100k,0.98) (SWaT 500k,0.99) (CrossTest,0.99)
};
\addlegendentry{Spectral}

\end{axis}
\end{tikzpicture}
\caption{Matthews Correlation Coefficient (MCC) comparison across datasets using Baseline, Community, and Spectral methods.}
\end{subfigure}

\caption{Detection capabilities comparison}

\label{fig:balanced_accuracy_vertical}
\end{figure}
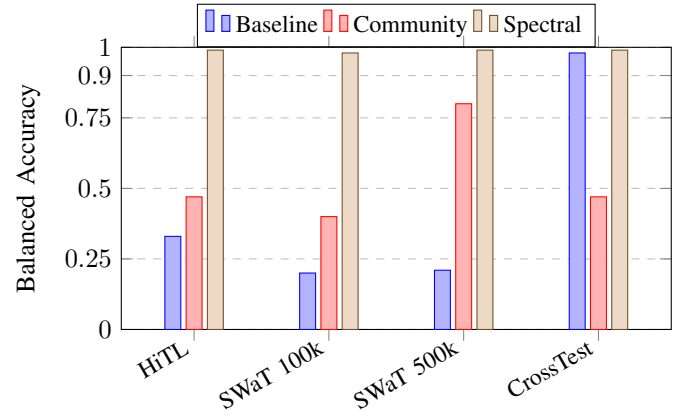
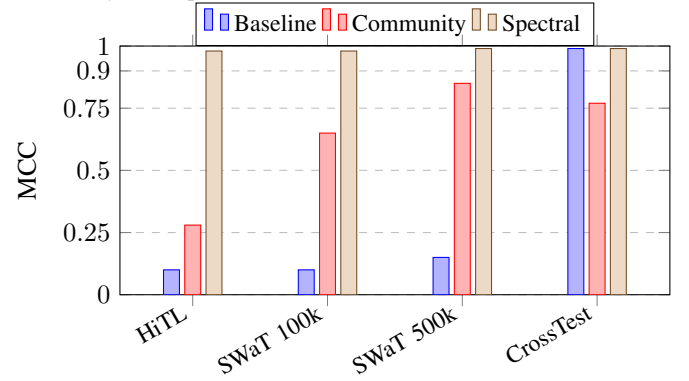

This section evaluates three approaches--- \textbf{Baseline}, which applies an XGBoost classifier to linear network traffic data; \textbf{Community metrics}, which extend the baseline features by incorporating community-based metrics; and \textbf{Spectral metrics}, which augment the baseline features with spectral metrics --- on multiclass cyberattack detection across the HiTL, SWaT 100k, SWaT 500k, and CrossTest datasets. The analysis relies on class-wise Precision, Recall, F1-Score, Balanced Accuracy, and Matthews Correlation Coefficient (MCC), computed from the predictions of the XGBoost classifier trained on the feature sets corresponding to the three evaluated approaches, as reported in Figures.~\ref{fig:model_comparison:precision},~\ref{fig:model_comparison:recall},~\ref{fig:model_comparison:f1score} and \ref{fig:balanced_accuracy_vertical}. All comparisons maintain identical dataset partitions and class definitions to ensure consistency across methods.

\begin{table}[t]
\centering
\caption{Dataset-level multiclass detection performance using Balanced Accuracy and MCC. Best results per dataset are shown in bold.}
\label{tab:performance_summary}
\resizebox{\columnwidth}{!}{
\begin{tabular}{l l c c}
\hline
Dataset & Approach & Balanced Accuracy & MCC \\
\hline

\multirow{3}{*}{HiTL}
 & Baseline  & 0.33 & 0.10 \\
 & Community & 0.47 & 0.28 \\
 & \textbf{Spectral} & \textbf{0.99} & \textbf{0.98} \\

\hline

\multirow{3}{*}{SWaT 100k}
 & Baseline  & 0.20 & 0.10 \\
 & Community & 0.40 & 0.65 \\
 & \textbf{Spectral} & \textbf{0.98} & \textbf{0.98} \\

\hline

\multirow{3}{*}{SWaT 500k}
 & Baseline  & 0.21 & 0.15 \\
 & Community & 0.80 & 0.85 \\
 & \textbf{Spectral} & \textbf{0.99} & \textbf{0.99} \\

\hline

\multirow{3}{*}{CrossTest}
 & Baseline  & 0.98 & 0.99 \\
 & Community & 0.47 & 0.77 \\
 & \textbf{Spectral} & \textbf{0.99} & \textbf{0.99} \\

\hline
\end{tabular}}
\end{table}

Table~\ref{tab:performance_summary} summarizes multiclass robustness across datasets. 
Baseline learning shows poor generalization on HiTL and SWaT datasets, with low Balanced Accuracy and MCC. 
Community metrics improve discrimination and correlation by incorporating structural graph information. 
Spectral metrics consistently achieve near-perfect performance across all datasets, demonstrating stable multiclass separability and scalability in heterogeneous cyberattack detection scenarios.

\begin{figure}[ht]
\centering
\begin{subfigure}[b]{0.4\textwidth}
    \centering
    \begin{tikzpicture}
    \begin{axis}[
        ylabel={Precision},
        xmin=0, xmax=2, ymin=0.5, ymax=1,
        xtick={0,1,2,3}, xticklabels={Benign, MitM, Physical Fault},
        xticklabel style={font=\tiny,rotate=45},
        ytick={0.5,0.75,0.9,1}, legend pos=south west,
        legend style={fill=none, font=\tiny}, 
        ymajorgrids=true, grid style=dashed,
        width=\textwidth, height=0.75\textwidth,
        legend style={
        at={(0.5,1.0)},
        anchor=south,
        legend columns=3
        }
    ]
    \addplot[color=red,mark=square] coordinates {
        (0,0.66)(1,0.63)(2,0.56)};
    \addlegendentry{Baseline}
    \addplot[color=blue,mark=square] coordinates {
        (0,0.71)(1,0.64)(2,0.71)};
    \addlegendentry{$cmetrics$}
    \addplot[color=orange,mark=square] coordinates {
        (0,0.99)(1,0.99)(2,0.98)};
    \addlegendentry{$smetrics$}
    \end{axis}
    \end{tikzpicture}
    \caption{HiTL Dataset}
    \label{fig:bot.f1score.1}
\end{subfigure}
\hspace{0.01\textwidth}
\begin{subfigure}[b]{0.4\textwidth}
    \centering
    \begin{tikzpicture}
    \begin{axis}[
        ylabel={Precision},
        xmin=0, xmax=4, ymin=0.0, ymax=1,
        xtick={0,1,2,3,4}, xticklabels={Benign, Disrupt Sensor, Exfiltration, Malware infection, Normal post attack},
        xticklabel style={font=\tiny,rotate=45},
        ytick={0.1,0.5,0.75,0.9,1}, legend pos=south west,
        legend style={fill=none, font=\tiny}, 
        ymajorgrids=true, grid style=dashed,
        width=\textwidth, height=0.75\textwidth,
        legend style={
        at={(0.5,1.0)},
        anchor=south,
        legend columns=3
        }
    ]
    \addplot[color=red,mark=square] coordinates {
        (0,0.72)(1,0.04)(2,0.28)(3,0.10)(4,0.11)};
    \addlegendentry{Baseline}
    \addplot[color=blue,mark=square] coordinates {
        (0,0.99)(1,0.32)(2,0.37)(3,0)(4,0.31)};
    \addlegendentry{$cmetrics$}
    \addplot[color=orange,mark=square] coordinates {
        (0,0.99)(1,0.98)(2,0.95)(3,0.99)(4,0.98)};
    \addlegendentry{$smetrics$}
    \end{axis}
    \end{tikzpicture}
    \caption{SWaT 100k Dataset}
    \label{fig:bot.f1score.2}
\end{subfigure}
\hspace{0.01\textwidth}
\begin{subfigure}[b]{0.4\textwidth}
    \centering
    \begin{tikzpicture}
    \begin{axis}[
        ylabel={Precision},
        xmin=0, xmax=4, ymin=0.1, ymax=1,
        xtick={0,1,2,3,4}, xticklabels={Benign, Disrupt Sensor, Exfiltration, Malware infection, Normal post attack},
        xticklabel style={font=\tiny,rotate=45},
        ytick={0.1,0.5,0.75,0.9,1}, legend pos=south west,
        legend style={fill=none, font=\tiny}, 
        ymajorgrids=true, grid style=dashed,
        width=\textwidth, height=0.75\textwidth,
        legend style={
        at={(0.5,1.0)},
        anchor=south,
        legend columns=3
        }
    ]
    \addplot[color=red,mark=square] coordinates {
        (0,0.17)(1,0.98)(2,0.13)(3,0.73)(4,0.19)};
    \addlegendentry{Baseline}
    \addplot[color=blue,mark=square] coordinates {
        (0,0.99)(1,0.50)(2,0.99)(3,0.99)(4,0.50)};
    \addlegendentry{$cmetrics$}
    \addplot[color=orange,mark=square] coordinates {
        (0,0.99)(1,0.99)(2,0.99)(3,0.99)(4,0.99)};
    \addlegendentry{$smetrics$}
    \end{axis}
    \end{tikzpicture}
    \caption{SWaT 500k Dataset}
    \label{fig:bot.f1score.3}
\end{subfigure}
\hspace{0.01\textwidth}
\begin{subfigure}[b]{0.4\textwidth}
    \centering
    \begin{tikzpicture}
    \begin{axis}[
        ylabel={Precision},
        xmin=0, xmax=2, ymin=0.8, ymax=1,
        xtick={0,1,2}, xticklabels={Benign, DoS, Scan},
        xticklabel style={font=\tiny,rotate=45},
        ytick={0.5,0.75,0.9,1}, legend pos=south west,
        legend style={fill=none, font=\tiny}, 
        ymajorgrids=true, grid style=dashed,
        width=\textwidth, height=0.75\textwidth,
        legend style={
        at={(0.5,1.0)},
        anchor=south,
        legend columns=3
        }
    ]
    \addplot[color=red,mark=square] coordinates {
        (0,0.84)(1,0.99)(2,0.93)};
    \addlegendentry{Baseline}
    \addplot[color=blue,mark=square] coordinates {
        (0,0.99)(1,0.97)(2,0.99)};
    \addlegendentry{$cmetrics$}
    \addplot[color=orange,mark=square] coordinates {
        (0,0.997)(1,0.99)(2,0.99)};
    \addlegendentry{$smetrics$}
    \end{axis}
    \end{tikzpicture}
    \caption{CrossTest Dataset}
    \label{fig:bot.f1score.4}
\end{subfigure}
\caption{Class-wise Precision results for three approaches across four datasets: HiTL, SWaT 100k, SWaT 500k, and CrossTest.}
\label{fig:model_comparison:precision}
\end{figure}

\begin{figure}[ht]
\centering
\begin{subfigure}[b]{0.4\textwidth}
    \centering
    \begin{tikzpicture}
    \begin{axis}[
        ylabel={Recall},
        xmin=0, xmax=2, ymin=0.0, ymax=1,
        xtick={0,1,2,3}, xticklabels={Benign, MitM, Physical Fault},
        xticklabel style={font=\tiny,rotate=45},
        ytick={0.0, 0.5,0.75,0.9,1}, legend pos=south west,
        legend style={fill=none, font=\tiny}, 
        ymajorgrids=true, grid style=dashed,
        width=\textwidth, height=0.75\textwidth,
        legend style={
        at={(0.5,1.0)},
        anchor=south,
        legend columns=3
        }
    ]
    \addplot[color=red,mark=square] coordinates {
        (0,0.99)(1,0.01)(2,0.00)};
    \addlegendentry{Baseline}
    \addplot[color=blue,mark=square] coordinates {
        (0,0.94)(1,0.25)(2,0.17)};
    \addlegendentry{$cmetrics$}
    \addplot[color=orange,mark=square] coordinates {
        (0,0.99)(1,0.98)(2,0.98)};
    \addlegendentry{$smetrics$}
    \end{axis}
    \end{tikzpicture}
    \caption{HiTL Dataset}
    \label{fig:bot.f1score.5}
\end{subfigure}
\hspace{0.02\textwidth}
\begin{subfigure}[b]{0.4\textwidth}
    \centering
    \begin{tikzpicture}
    \begin{axis}[
        ylabel={Recall},
        xmin=0, xmax=4, ymin=0.0, ymax=1,
        xtick={0,1,2,3,4}, xticklabels={Benign, Disrupt Sensor, Exfiltration, Malware infection, Normal post attack},
        xticklabel style={font=\tiny,rotate=45},
        ytick={0.1,0.5,0.75,0.9,1}, legend pos=south west,
        legend style={fill=none, font=\tiny}, 
        ymajorgrids=true, grid style=dashed,
        width=\textwidth, height=0.75\textwidth,
        legend style={
        at={(0.5,1.0)},
        anchor=south,
        legend columns=3
        }
    ]
    \addplot[color=red,mark=square] coordinates {
        (0,0.99)(1,0.001)(2,0.006)(3,0.002)(4,0.003)};
    \addlegendentry{Baseline}
    \addplot[color=blue,mark=square] coordinates {
        (0,0.99)(1,0.007)(2,0.98)(3,001)(4,0.01)};
    \addlegendentry{$cmetrics$}
    \addplot[color=orange,mark=square] coordinates {
        (0,0.99)(1,0.97)(2,0.99)(3,0.98)(4,0.98)};
    \addlegendentry{$smetrics$}
    \end{axis}
    \end{tikzpicture}
    \caption{SWaT 100k Dataset}
    \label{fig:bot.f1score.6}
\end{subfigure}
\hspace{0.02\textwidth}
\begin{subfigure}[b]{0.4\textwidth}
    \centering
    \begin{tikzpicture}
    \begin{axis}[
        ylabel={Recall},
        xmin=0, xmax=4, ymin=0.0, ymax=1,
        xtick={0,1,2,3,4}, xticklabels={Benign, Disrupt Sensor, Exfiltration, Malware infection, Normal post attack},
        xticklabel style={font=\tiny,rotate=45},
        ytick={0.1, 0.5,0.75,0.9,1}, legend pos=south west,
        legend style={fill=none, font=\tiny}, 
        ymajorgrids=true, grid style=dashed,
        width=\textwidth, height=0.75\textwidth,
        legend style={
        at={(0.5,1.0)},
        anchor=south,
        legend columns=3
        }
    ]
    \addplot[color=red,mark=square] coordinates {
        (0,0.99)(1,0.00)(2,0.08)(3,0.00)(4,0.00)};
    \addlegendentry{Baseline}
    \addplot[color=blue,mark=square] coordinates {
        (0,0.99)(1,0.65)(2,0.99)(3,0.99)(4,0.35)};
    \addlegendentry{$cmetrics$}
    \addplot[color=orange,mark=square] coordinates {
        (0,0.99)(1,0.99)(2,0.99)(3,0.99)(4,0.99)};
    \addlegendentry{$smetrics$}
    \end{axis}
    \end{tikzpicture}
    \caption{SWaT 500k Dataset}
    \label{fig:bot.f1score.7}
\end{subfigure}
\hspace{0.02\textwidth}
\begin{subfigure}[b]{0.4\textwidth}
    \centering
    \begin{tikzpicture}
    \begin{axis}[
        ylabel={Recall},
        xmin=0, xmax=2, ymin=0.9, ymax=1,
        xtick={0,1,2}, xticklabels={Benign, DoS, Scan},
        xticklabel style={font=\tiny,rotate=45},
        ytick={0.9,0.95,0.99,1}, legend pos=south west,
        legend style={fill=none, font=\tiny}, 
        ymajorgrids=true, grid style=dashed,
        width=\textwidth, height=0.75\textwidth,
        legend style={
        at={(0.5,1.0)},
        anchor=south,
        legend columns=3
        }
    ]
    \addplot[color=red,mark=square] coordinates {
        (0,0.998)(1,0.998)(2,0.994)};
    \addlegendentry{Baseline}
    \addplot[color=blue,mark=square] coordinates {
        (0,0.97)(1,0.00)(2,0.92)};
    \addlegendentry{$cmetrics$}
    \addplot[color=orange,mark=square] coordinates {
        (0,0.999)(1,0.991)(2,0.993)};
    \addlegendentry{$smetrics$}
    \end{axis}
    \end{tikzpicture}
    \caption{CrossTest Dataset}
    \label{fig:bot.f1score.8}
\end{subfigure}
\caption{Class-wise Recall results for three approaches across the HiTL, SWaT 100k, SWaT 500k, and CrossTest datasets.}
\label{fig:model_comparison:recall}
\end{figure}

\section{Conclusions}
\label{sec: conc}
This work presented a demonstration of how topology-driven approaches can be used for detection cyberattacks in water distribution ICS. With the help of GPML library, the network traffic was modeled into dynamic graphs which allows to analyze both local interaction in the network and global structure changes induced by cyber and physical attack. Graph community metrics provided the temporal evaluation of the graphs while spectral metrics helps in analyzing the changes in the topology in a time window. The dynamic community and spectral analysis methodolgies has been applied to three water distribution datasets, HITL, SWaT, and CrossTest. These datasets consist of different physical and cyber attacks. From the results, it can be deduced that both spectral and community metrics achieve higher precision, F1-score, and recall compared to the baseline machine learning model.

\bibliographystyle{alpha}
\bibliography{mct}

\end{document}